\documentstyle[preprint,aps,prabib]{revtex}

\begin{document}

\title{Orbital and field angular momentum in the nucleon}

\author{D. Singleton
\thanks{E-mail address : dougs@csufresno.edu}}
\address{Dept. of Physics, CSU Fresno, 2345 East San Ramon Ave.
M/S 37, Fresno, CA 93740-8031}
\author{V. Dzhunushaliev
\thanks{E-mail address: bars@krsu.edu.kg}}
\address{Theor. Physics Dept.,
Kyrgyz State National University,
720024, Bishkek, Kyrgyzstan, and Universit{\"a}t Potsdam, Institut
f{\"u}r Mathematik, 14469, Potsdam, Germany}

\date{\today}

\maketitle

\begin{abstract}
The nucleon spin problem raises
experimental and theoretical questions regarding the
contribution of the orbital angular momentum of the quarks
to the total spin of the nucleon. In this article we
examine the commutation relationships of various
operators that contribute to the total
angular momentum of the nucleon. We find that
the sum of the {\em orbital plus gluon field angular momenta}
should satisfy the angular momentum commutators, at least
up to the one-loop level. This requirement on the sum of these
operators imposes a non-trivial 
restriction on the form of the color electric and magnetic
fields. This is similar to the magnetic monopole/electric
charge system where it is only the sum of the orbital
plus field angular momentum that satisfies the correct
commutation relationships.

\end{abstract}
\narrowtext

{\flushleft{\big . \bf{1. INTRODUCTION}}}

Since the European Muon Collaboration (EMC) experiment
at CERN where muons were scattered off polarized protons 
\cite{emc} it has been realized that,
contrary to the simple quark model, the spin of the nucleon comes
not only from the spin of the valence quarks, but also has 
contributions from the quark orbital angular momentum and 
from the gluons. Mathematically this is written as
\begin{equation}
\label{1}
{1 \over 2} = {1 \over 2} \Delta \Sigma (q^2) + L_q(q^2) + J_g (q^2)
\end{equation}
where $q^2$ is the scale at which the operators are determined.
The spin $1/2$ of the nucleon is broken into contributions
from the quark spin (${1 \over 2} \Delta \Sigma(q^2)$), the quark orbital
angular momentum ($L_q (q^2)$), and the total gluon contribution ($J_g (q^2)$).
(Technically one should also include the contribution of the photons,
$J_{\gamma}$ \cite{d1} and weak gauge bosons $J_{WZ}$). In Ref. \cite{ji}
an explicitly gauge-invariant form of the total QCD angular
momentum was given
\begin{equation}
\label{2}
{\vec J_{QCD}} = \int d^3 x \left[ {1 \over 2} {\bar \psi} {\vec \gamma}
\gamma _5 \psi + \psi ^{\dag} ({\vec x} \times (-i {\vec D})) \psi
+ {\vec x} \times ({\vec E}^a \times {\vec B}^a) \right]
\end{equation}
where $D_i = \partial _i - i g A_i ^a T^a$ is the covariant derivative
with $T^a$ the $SU(3)$ generators. The first term in Eq. (\ref{2})
is associated with the spin of the quarks, the second term with the
orbital angular momentum of the quarks, and the last term is the
total gluonic contribution. Eq. (\ref{2}) can be cast in four-vector
notation by defining
\begin{equation}
\label{2a}
J^{\mu \nu} = \int d^3 x M^{0 \mu \nu} ({\vec x})
\end{equation}
where $M^{\alpha \mu \nu}$ is a rank 3 tensor which can be defined
in terms of the energy-momentum tensor, $T^{\mu \nu}$
\begin{equation}
\label{2b}
M^{\alpha \mu \nu} = T^{\alpha \nu} x^{\mu} - T^{\alpha \mu} x^{\nu}
\end{equation}
The QCD energy momentum tensor is given by
\begin{equation}
\label{2c}
T^{\mu \nu} = {1 \over 4} {\bar \psi} \gamma ^{(\mu}
i \stackrel{\leftrightarrow}{D} \; ^{\nu )} \psi +
\left( {1 \over 4} g^{\mu \nu} F^{\alpha \beta a}  F_{\alpha \beta} ^a
- F^{\mu \alpha a} F ^{\nu a} _{\alpha} \right)
\end{equation}
where $\gamma ^{(\mu} i \stackrel{\leftrightarrow}{D} \; ^{\nu )}$
means that the indices are symmetrized, and $a$ is a group index.
With these definitions the angular momentum
components of Eq. (\ref{2}) can be expressed in terms of
Eq. (\ref{2a}) as $J^i _{QCD} = {1 \over 2} \epsilon ^{ijk} J^{jk}$.
In examining the commutators of the various terms in Eq. (\ref{2})
we will deal with the angular momentum density operators. Also
for the first two terms in Eq. (\ref{2}) we will not always write
out the $\psi$'s.

Conventionally a quantum angular momentum operator should satisfy
the standard angular momentum commutation relationship
\begin{equation}
\label{3}
[J_i , J_j ] = i \epsilon_{ijk} J_k
\end{equation}
For example the first term (${1 \over 2} \gamma _0 {\vec \gamma}
\gamma_5$) in Eq. (\ref{2}) can be written as
\begin{equation}
\label{4}
{1 \over 2} \left(
\begin{array}{cc}
{\vec \sigma} & 0 \\
0 & {\vec \sigma}
\end{array}
\right)
\end{equation}
Since the ${\vec \sigma}$ matrices satisfy $[\sigma _i , \sigma _j]
=2 i \epsilon _{ijk} \sigma _k$ one finds that
the first term in Eq. (\ref{2}) satisfies Eq. (\ref{3}).
At higher orders the various operators in Eq. (\ref{2})
can be scale dependent and mix with one another,
which complicates matters. However, in Ref. \cite{ji1}
it was shown that at the one-loop level the contribution,
${1 \over 2} \Delta \Sigma$, associated with the first term in Eq. (\ref{2}),
was scale independent so that, at least up to this order, the
first operator in Eq. (\ref{2}) should satisfy Eq. (\ref{3}).
The quantity, ${\vec J}_{QCD}$, should always
satisfy Eq. (\ref{3}) regardless of $q^2$, since it is
the total, conserved angular momentum of the nucleon.
If we rearrange Eq. (\ref{2}) by placing the spin term
on the left hand side (so that ${\vec J}_{QCD} - {\vec S} =
{\vec L}_{QCD} +{\vec G}_{QCD}$) then at least to the one loop
level ${\vec J}_{QCD} - {\vec S}$, and therefore
${\vec L}_{QCD} +{\vec G}_{QCD}$, should satisfy Eq. (\ref{3}).
We will find that this requirement places restrictions on
the color electric/magnetic fields.  

{\flushleft {\big . \bf{2. QUARK ORBITAL ANGULAR MOMENTUM}}} 

The quark orbital angular momentum part of Eq. (\ref{2}) can be written
in component form as $L_{QCD} ^i = - i\epsilon^{ilm} x^l D^m$
with $D^m = \partial ^m - i g A^{ma} T^a$. Inserting this into
Eq. (\ref{3}) gives after some commutator algebra
\begin{equation}
\label{5}
[L_{QCD} ^i , L_{QCD} ^j] = -\epsilon ^{ilm} \epsilon ^{jpq}
\left( x^l x^p [D^m, D^q] + x^l [D^m, x^p] D^q 
+x^p [x^l , D^q] D^m \right)
\end{equation}
Using $[D^m , x^p] = \delta ^{mp}$ and $[D^m , D^q] = i g
G^{mq} = i g \epsilon ^{mqk} B^k$ (where the matrix $B^k = B^{ak} T^a$ is
the color magnetic field) we can transform Eq. (\ref{5}) to
\begin{equation}
\label{6}
[L_{QCD} ^i , L_{QCD} ^j] = -\epsilon^{ilm} \left( \epsilon ^{jmq}
x^l D^q - \epsilon ^{jpl} x^p D^m + i g x^l x^p \epsilon ^{jpq}
\epsilon ^{mqk} B^k \right)
\end{equation}
Using the standard formula $\epsilon^{ilm} \epsilon^{jqm} =
\delta ^{ij} \delta ^{lq} - \delta ^{iq} \delta ^{lj}$ several
times and eliminating terms like $x^l x^m \epsilon ^{ilm}$ by
symmetry we find
\begin{equation}
\label{7}
[L_{QCD} ^i , L_{QCD} ^j] = \left( x^i D^j - x^j D^i + i g 
\epsilon ^{ilj} x^l x^k B^k \right)
\end{equation}
The first two terms can be written as $x^i D^j - x^j D^i =
i \epsilon ^{ijk} \epsilon ^{klm} x^l (-i D^m) = i
\epsilon ^{ijk} L ^k _{QCD}$. Using this our final result
is
\begin{equation}
\label{8}
[L^i _{QCD} , L^j _{QCD}] = i \epsilon ^{ijk} L^k _{QCD}
- i \epsilon ^{ijk} (g x^k x^l B^l)
\end{equation}
where in the last term we have renamed some of the dummy indices.
The second term in Eq. (\ref{8}) prevents ${\vec L}_{QCD}$ from
satisfying Eq. (\ref{3}). This has also been remarked
upon in Ref. \cite{wang}. One could force ${\vec L}_{QCD}$
to satisfy Eq. (\ref{3}) by requiring that color magnetic
fields obey the condition $x^l B^l = 0$. This would
be a strong restriction on the form of the color magnetic
fields. This option is probably not viable, since some 
successful phenomenological models of baryons
\cite{georgi} (or see Ref. \cite{don} for a general overview)
explain the mass differences between baryons of similar quark 
content but different spins in terms of a color magnetic
dipole-dipole interaction, in analogy with electromagnetism.
For such a color magnetic dipole field $x^l B^l \ne 0$.
In addition there is no real reason to try and force
${\vec L}_{QCD}$ to satisfy Eq. (\ref{3}), since from the discussion
at the end of the previous section it is only the sum, ${\vec L}_{QCD}
+{\vec G}_{QCD} = {\vec J}_{QCD} - {\vec S}$, which
was required to satisfy Eq. (\ref{3}), at least up to the
one-loop level.

Such a situation, where it is only the sum of the field plus
orbital angular momentum which satisfies Eq. (\ref{3}),
is encountered in the electromagnetic system of an 
electric charge and magnetic monopole \cite{peskin}.
Also if one considers an electric charge/point magnetic dipole
system one again finds \cite{das} that it is only
the combination of orbital plus field angular momentum that
satisfies the angular momentum commutators. In the next section
we make some assumptions about the form of the color electric and
color magnetic fields of the nucleon, and show that the angular
momentum commutators impose a non-trivial restriction on the form of
these fields. In the electromagnetic examples cited the fields
are known explicitly and one must only check that
the resulting particle plus field angular momentum satisfies
Eq. (\ref{3}). In the QCD case, however, the explicit
form of the color fields is not known, so this restriction coming
from the angular momentum commutation relationships might give
some extra insight into the structure of these fields. 

{\flushleft {\big . \bf{3. FIELD ANGULAR MOMENTUM}}} 

At the one-loop level the combination ${\vec J}_{QCD} - {\vec S}$
obeys the angular momentum commutation rules, since each
term separately obeys them (${\vec J}_{QCD}$ satisfies
the angular momentum commutation rules to all orders). Thus
at the same one-loop level ${\vec L}_{QCD} +{\vec G}_{QCD} =
{\vec J}_{QCD} - {\vec S}$ is also required to satisfy the
angular momentum commutation relationships.
Applying this requirement to the combination of
orbital plus gluon field angular momentum
we will find that certain restrictions
are placed on the form of the fields. The commutator
of the orbital plus field angular momentum operators is
\begin{eqnarray}
\label{9}
[(L_{QCD}^i + G_{QCD} ^i) , (L_{QCD} ^j + G_{QCD}^j)] &=&
[L_{QCD} ^i , L_{QCD} ^j] + [G_{QCD} ^i , G_{QCD} ^j]
\nonumber \\
&+& [G_{QCD} ^i , L_{QCD} ^j] + [L_{QCD} ^i , G_{QCD} ^j]
\end{eqnarray}
The first term on the right hand side was calculated in
the last section. In order to calculate the last three
terms we need to express the field angular momentum in
index notation as $G ^i _{QCD} = \epsilon ^{ijk} x^j
\epsilon ^{klm} E_l ^a B_m ^a$. Next, the matrix chromoelectric
and chromomagnetic fields are usually defined as
${\cal E} _p (x) = E^a _p (x) T^a$
and ${\cal B} _p (x) = B^a _p (x) T^a$. Using the standard
normalization for the ${T^a} 's$ ($Tr [T^a T^b] = \delta ^{ab} /2$
where $Tr$ is the trace) gives
\begin{equation}
\label{10}
E^a _p (x) = 2 Tr [T^a {\cal E} _p (x)] \; \; \; \;  \; \; \; \;
B^a _p (x) = 2 Tr [T^a {\cal B} _p (x)]
\end{equation}
An example of specific forms for the chromoelectric and
chromomagnetic fields in a generic quark model can be found
in Ref. \cite{jaffe}. With Eq. (\ref{10}) the second commutator in
Eq. (\ref{9}) becomes
\begin{eqnarray}
\label{11}
[G_{QCD}^i , G_{QCD}^j] &=& [\epsilon ^{ilm} \epsilon ^{mpq} x^l
E_p ^a B_q ^a , \epsilon ^{jrs} \epsilon ^{stu} x^r E_t ^b B_u ^b ]
\nonumber \\
&=& 16 \epsilon ^{ilm} \epsilon ^{mpq} \epsilon ^{jrs} \epsilon ^{stu}
x^l x^r  [Tr(T^a {\cal E}_p) Tr (T^a {\cal B} _q)  ,
Tr (T^b {\cal E}_t) Tr(T^b {\cal B} _u)]
\end{eqnarray}
Now for any matrices $A, B, C, D$
\begin{equation}
[Tr(A) Tr(B) , Tr(C) Tr (D)] = 0
\end{equation}
Thus the right hand side of Eq. (\ref{11}) equals zero,
and $[G_{QCD} ^i , G_{QCD} ^j] = 0$ with respect to the
color group structure of $G_{QCD} ^i$. However,
$[G_{QCD} ^i , G_{QCD} ^j]$ could have a possible non-zero
contribution arising from the canonical commutation relationships
of the non-Abelian gauge potentials, $A_{\mu} ^a$. First,
using standard commutator techniques the complex
commutator of Eq. (\ref{11}) can be simplified into a sum of
two term commutators
\begin{equation}
\label{11a}
[E_p ^a B_q ^a , E_t ^b B_u ^b] =  [E_p ^a , E_t ^b] B_u ^b B_q^a
+ E_p ^a E_t ^b [B_q ^a , B_u ^b] 
+ E_t ^b [E_p^a , B_u ^b] B_q^a + E_p ^a [B_q^a , E_t ^b] B_u^b 
\end{equation}
In order to evaluate these commutators we expand the
chromoelectric and chromomagnetic fields in terms of
the gauge potentials. The general expressions are
\begin{eqnarray}
\label{11b}
E_p ^a &=& F_{0 p} ^a = \partial _0 A_p ^a -\partial _p A_0 ^a +
g f^{abc} A^b _0 A^c _p \nonumber \\
B_p ^a &=& F_{m n} ^a = \partial _m A_n ^a -\partial _n A_m ^a +
g f^{abc} A^b _m A^c _n
\end{eqnarray}
where $f^{abc}$ are the group structure constants.
In the canonical formalism the
gauge potentials satisfy the following commutation rules \cite{ms}
\begin{eqnarray}
\label{ccr}
&[&\partial _0 A_{\mu} ^a ({\bf x} , t) , A_{\nu} ^b ({\bf x}' , t)]
= i g_{\mu \nu} \delta ^{ab} \delta ^3 ({\bf x} - {\bf x}')
\nonumber \\
&[&A_{\mu} ^a ({\bf x} , t) , A_{\nu} ^a ({\bf x}' , t)] =
[\partial _0 A_{\mu} ^a ({\bf x} , t) , \partial _0
A_{\nu} ^a ({\bf x}' , t)] =  0
\end{eqnarray}
These are similar to the Abelian field commutators except for the
Kronecker delta, $\delta ^{ab}$, in the group indices.
These commutation relationships along with the expansions given
in Eq. (\ref{11b}) give rise to the possibility that some of the
commutators in Eq. (\ref{11a}) may be non-zero. The
expansion of the chromomagnetic field in Eq. (\ref{11b})
involves only $A_n ^a$ and {\em spatial} derivatives of
$A_n ^a$, so the second commutator of Eq. (\ref{ccr}) implies that
$[B_q ^a , B_u ^b] = 0$ in Eq. (\ref{11a}). The other commutators
of Eq. (\ref{11a}) involve the chromoelectric field, which in the
most general case has $\partial _0 A_p ^a$ terms. Thus because
of the non-zero first commutator in Eq. (\ref{ccr}) the remaining
three terms from Eq. (\ref{11a}) could in general be non-zero.
For example, in the Abelian
case a similar development using the {\it time-dependent} electric and
magnetic fields of a photon results in the commutators between
the electric and magnetic fields ({\it i.e.} $[E_i , B_j]$) being
non-zero. These non-zero commutators between $E_i$ and $B_j$ ensure
that the photon angular momentum operator satisfies Eq.
(\ref{3}) \cite{merz}. Also in the general non-Abelian case
one would expect some or all of the commutators from Eq.
(\ref{11a}) to be non-zero. However, since the nucleon is
a bound state we will assume that the gauge fields are
static -- $\partial _0 A_{\mu} ^a = 0$ -- so that
the chromoelectric field for this specific case of the nucleon
simplifies to
\begin{equation}
E_p ^a = -\partial _p A_0 ^a + g f^{abc} A^b _0 A^c _p 
\end{equation}
Under this assumption the remaining commutators of Eq. (\ref{11a})
involve only the non-Abelian gauge potentials, $A_{\mu} ^a$,
or their {\it spatial} derivatives. As with the pure chromomagnetic
commutator, $[B_q ^a , B_u ^b]$, the second commutator
from Eq. (\ref{ccr}) now also implies that the
remaining commutators from Eq. (\ref{11a}) are zero. This in turn
gives $[G_{QCD} ^i , G_{QCD} ^j] = 0$. One subtle point in the
above considerations is that rigorously one should impose
the time-independent condition not on the gauge field
operators, $A_{\mu} ^a$, via $\partial _0 A_{\mu} ^a = 0$, but
rather one should impose this as a condition on the nucleon
bound states. Taking $\vert p \frac{1}{2} \rangle$  as the state
vector for the nucleon bound state of a certain momentum and
helicity, the time-independent condition becomes
$\partial _0 A_{\mu} ^a \vert p \frac{1}{2} \rangle= 0$.
Using this one can show that each term on the
right hand side of Eq. (\ref{11a}) vanishes.
For example, taking the third term and inserting
$1 = \sum \vert p \frac{1}{2} \rangle
\langle p \frac{1}{2} \vert$ or
$1 = \sum_{'} \vert p' \frac{1}{2} \; ' \rangle
\langle p' \frac{1}{2} \; ' \vert$ appropriately gives
\begin{equation}
\label{12}
\sum \sum_{'}
\left . \left . E_t ^b \right\vert p \frac{1}{2}
\right\rangle \left\langle p \frac{1}{2} \left\vert
[\partial _0 A_p^a , B_u ^b ] \right\vert p' \frac{1}{2} \; ' \right\rangle
\left\langle p' \frac{1}{2} \; ' \left\vert B_q ^a  \right . \right .
\end{equation}
Only the time derivative part of the chromoelectric field from
inside the commutator has been written out, since by Eq. (\ref{ccr})
this is the only term which could give a non-zero result in
the commutator. Expanding the commutator part of Eq. (\ref{12})
and again inserting $1 = \sum_{''} \vert p '' \frac{1}{2}\; '' \rangle
\langle p'' \frac{1}{2}\; '' \vert$ appropriately gives
\begin{equation}
\sum_{''} \left( \left\langle p \frac{1}{2} 
\left\vert \partial _0 A_p^a
\right\vert p'' \frac{1}{2} \; '' \right\rangle
\left\langle p'' \frac{1}{2} \; '' \left\vert B_u ^b
\right\vert p' \frac{1}{2} \; ' \right\rangle
-
\left\langle p \frac{1}{2}  \left\vert B_u ^b
\right\vert p'' \frac{1}{2} \; '' \right\rangle
\left\langle p'' \frac{1}{2} \; '' \left\vert \partial _0 A_p^a
\right\vert p' \frac{1}{2} \; ' \right\rangle \right)
\end{equation}
Each term above separately equals zero by the condition
$\partial _0 A_{\mu} ^a \vert p \frac{1}{2} \rangle=
\partial _0 A_{\mu} ^a \vert p' \frac{1}{2} \; ' \rangle=
... = 0$. In the same way all the other terms from Eq. (\ref{11a})
which contain the possibly non-trivial $\partial _0 A_p ^a$ factors
in the commutator can be shown to vanish by appropriately inserting
$1= \sum \vert p \frac{1}{2} \rangle \langle p \frac{1}{2} \vert$
and imposing the condition $\partial _0 A_{\mu} ^a \vert p
\frac{1}{2} \rangle = \partial _0 A_{\mu} ^a \vert p'
\frac{1}{2} \; ' \rangle = ... = 0$.

The result that $[G_{QCD} ^i , G_{QCD} ^j] = 0$
may seem strange, but a similar result arises in the electromagnetic
system of a charge and monopole at rest with respect to one another.
The commutator of the components of the electromagnetic field
angular momentum ($G_{EM} ^i$) among themselves could be calculated
by expanding the electric and magnetic fields in terms of the
Abelian gauge potential and using equations Eqs. (\ref{11})
(\ref{ccr}) with the color indices dropped. The only non-trivial
terms would come from commutators between $\partial ^0 A^i$ and
$A^j$ or $\partial ^j A^k$. However, since the electric and
magnetic charges are at rest, the gauge fields are time-independent
so that $\partial ^0 A^i = 0$, and one again finds that the commutator
for the components of the electromagnetic field angular momentum
with itself are zero. In this electromagnetic case the calculation
is actually much easier if one first calculates the explicit form
for the field angular momentum, namely $G_{EM} ^i = e g x^i / r$
\cite{jackson} where $e , g, x^i$ are the electric charge, magnetic
charge and displacement between the two charges respectively. The
commutator of the components of $G_{EM} ^i$ among themselves
is zero since $x^i / r$ commutes with itself. In the color field
case this latter procedure was not an option since, unlike the
Abelian charge/monopole case, one does not have explicit forms for
the chromoelectric and chromomagnetic field, so that one can not
obtain an explicit form for $G_{QCD} ^i$.

In the present context the fact that
${\vec G}_{QCD}$ does not satisfy Eq.({\ref{3}) is not crucial.
If on the other hand one 
considered pure gluon states (glueballs) -
so that ${\vec J}_{QCD} = {\vec G}_{QCD}$ - then the
gluon field angular momentum {\it would be} the total, conserved
angular momentum. In this case ${\vec G}_{QCD}$ must satisfy Eq. (\ref{3}).
The fact that we get $[G_{QCD} ^i , G_{QCD} ^j] = 0$ could have
several possible resolutions : First, the form
that we used from the color fields in Eq. (\ref{10})
may not be valid for pure gluon states.
One might have to consider functions (${\cal E} _p (x) ,{\cal B} _p (x)$)
which have a non-trivial commutation with one another beyond those
already coming form the group factors and/or the annihilation and
creation operator representation of the gauge fields.
Also one could question the ability to split the
spatial and group factors as in Eq. (\ref{10}).
Such a change in the form of the non-Abelian electric and magnetic
fields would require a similar change in the analogous expression
for the matrix gauge potentials, $A_{\mu} = T^a A^a _{\mu}$. Since this
form of the potential is central in arriving at various
fundamental relationships of Yang-Mills theory ({\it e.g.}
$[D^{\mu} , D^{\nu}] = i g F^{\mu \nu}$) it is hard to see
how one could change Eq. (\ref{10}) without significantly
affecting the structure of Yang-Mills theory.
A second possible resolution to having
$[G^i _{QCD} , G^j _{QCD}] =0$ would be if pure gluon
systems always had zero angular momentum ({\it i.e.} if glueballs
were restricted to being spin zero objects). One might be tempted
to immediately discard this conclusion since one expects by
simple angular momentum addition that a glueball with two valence
gluons should have angular momenta of $0, 2 ...$, and for
a glueball with three valence gluons one could have in addition
odd angular momenta ($1, 3, ...$). However, such considerations
are based on the same type of simple arguments that had the
spin of the nucleon coming from the spin of the valence
quarks. Since the EMC experiments have shown that such a simple
picture is not correct for quark bound states like the nucleon
it is not unreasonable to raise questions about applying the
same procedure to pure gluon bound states.

The third term in Eq. (\ref{9}) can be written as
\begin{eqnarray}
\label{13}
[G_{QCD} ^i , L_{QCD} ^j ] &=& [\epsilon ^{ilm} \epsilon ^{mpq}
x^l E^{pa} B^{qa} , \epsilon ^{jrs} x^r (-i D^s) ] 
\nonumber \\
&=& -i \epsilon ^{ilm} \epsilon ^{mpq} \epsilon ^{jrs} \left(
x^l x^r [E^{pa} B^{qa} , D^s] + x^r [x^l, D^s] E^{pa} B^{qa} 
\right)
\end{eqnarray}
The first commutator can be broken into two parts as $[E^{pa} B^{qa} ,
\partial ^s] - i g [E^{pa} B^{qa} , A^{sb} T^b]$. The last part
can be reduced to $-i g [4 Tr(T^a {\cal E}^p) Tr(T^a {\cal B}^q) ,
T^b A^{sb} ]$ (the $T^b A^{sb}$ matrix term should be bracketed by
${\psi}^{\dag}$ and $\psi$). This commutator could be non-trivial from either
the group factor or from representing $A^{sb}$ (and therefore ${\cal E}^p ,
{\cal B}^q$) in terms of creation/annihilation operators.
The group factors do not give a non-zero commutator
since the first term is a trace over the group factors.
Representing the gauge fields
as operators also does not give a non-zero contribution to the
commutator for the same reason as for the pure gluon commutator :
only $[\partial _0 A_{\mu} ^a ({\bf x} , t) , A_{\nu} ^a
({\bf x}' , t)]$ is non-zero, and in the present case
we are assuming static fields so $\partial _0 A_{\mu} ^a ({\bf x} , t)
=0$ (or more rigorously $\partial _0 A_{\mu} ^a ({\bf x} , t)
\vert p \frac{1}{2} \rangle = 0$).

The first part becomes $[E^{pa} B^{qa} , \partial ^s] =
- \partial ^s (E^{pa} B^{qa})$. The second commutator in Eq. (\ref{13})
is just $[x^l , D^s] = -\delta ^{ls}$. Combining these
results and using $\epsilon^{iml} \epsilon ^{jrl} =
\delta ^{ij} \delta ^{mr} - \delta ^{ir} \delta ^{mj}$ we find
\begin{eqnarray}
\label{14}
[G_{QCD} ^i , L_{QCD} ^j] &=& i \left( x^i \epsilon ^{jpq} E^{pa} B^{qa}
- \delta ^{ij} \epsilon ^{mpq} x^m E^{pa} B^{qa} \right)
\nonumber \\
&+& i \epsilon ^{jrs} x^l x^r \partial ^s \left( E^{ia} B^{la}
- E^{la} B^{ia} \right)
\end{eqnarray}
The other cross term from Eq. (\ref{9}) can be obtained from 
Eq. (\ref{14}) by multiplying the result by a minus sign to account for
the reversed order of the terms, and by interchanging the indices
$i \leftrightarrow j$. Combining these two cross terms of Eq. (\ref{9})
gives
\begin{eqnarray}
\label{15}
[G_{QCD}^i , L_{QCD} ^j] + [L_{QCD}^i , G_{QCD} ^j] &=&
 i \left( x^i \epsilon ^{jpq} E^{pa} B^{qa} - x^j
\epsilon ^{ipq} E^{pa} B^{qa} \right) 
\nonumber \\
&+& i \epsilon ^{jrs} x^l x^r \partial ^s \left( E^{ia} B^{la}
- E^{la} B^{ia} \right)
\nonumber \\
&+&i \epsilon ^{irs} x^l x^r \partial ^s \left( E^{la} B^{ja} -
E ^{ja} B^{la} \right)
\end{eqnarray}
The first term on the right hand side becomes
$i \epsilon ^{ijk} (\epsilon ^{klm} x^l \epsilon ^{mpq}
E^{pa} B^{qa}) = i \epsilon ^{ijk} G^k _{QCD}$. Also
by making the definitions $A^{ij} \equiv \epsilon ^{irs}
x^l x^r \partial ^s (E^{ja} B^{la} )$ and $C^{ij} \equiv
\epsilon ^{irs} x^l x^r \partial ^s (E^{la} B^{ja})$ we
can write the right hand side of Eq. (\ref{15}) as
\begin{equation}
\label{16}
i \epsilon ^{ijk} G^k _{QCD} + i (A^{ji} - A^{ij})
+i(C^{ij} - C^{ji}) 
 =
i \epsilon ^{ijk} G^k _{QCD} + i \epsilon ^{ijk}
\left( \epsilon ^{kmn} C^{mn} - \epsilon ^{kmn} A^{mn} \right)
\end{equation}
Using $\epsilon ^{kmn} \epsilon ^{mrs} = - \delta ^{kr} \delta ^{ns}
+ \delta ^{ks} \delta ^{nr}$ and renaming some summed over dummy
indices Eq. (\ref{16}) becomes
\begin{equation}
\label{17}
i \epsilon ^{ijk} G^k _{QCD} + i \epsilon ^{ijk} x^l x^k 
\partial ^s \left( E^{sa} B^{la} - E^{la} B^{sa} \right) 
\end{equation}
Combining the result of Eq. (\ref{17}) with the results from
Eqs. (\ref{8}) and (\ref{11}) we find that the commutator for
the combined field plus orbital angular momentum density is
\begin{eqnarray}
\label{18}
[L^i _{QCD} + G^i _{QCD} , L^j _{QCD} + G^j _{QCD}] &=&
i \epsilon ^{ijk} \left( L^k _{QCD} + G^k _{QCD} \right)
\nonumber \\
&+& i \epsilon ^{ijk} x^l x^k \left( - g B^{la} T^a +
\partial ^s \left( E^{sa} B^{la} - E^{la} B^{sa} \right)
\right)
\end{eqnarray}
where the $L_{QCD} ^k$ and $g B^{la} T^a$ terms
should be bracketed by $\psi ^{\dag}$ and $\psi$.
The first line on the right hand side is the correct term to
close the angular momentum algebra of the combination
$L^i _{QCD} + G^i _{QCD}$, however the second line prevents
this. Now the combination $L^i _{QCD} + G^i _{QCD}$
{\em would} satisfy Eq. (\ref{3}) if the color electric and
magnetic fields, and/or the spinors,
$\psi$, obeyed certain restrictions. For example, if
\begin{equation}
\label{19}
- g x^l {\bar \psi} B^{la} T^a \psi 
+ x^l \partial ^s \left( E^{sa} B^{la} - E^{la} B^{sa} \right) = 0
\end{equation}
or in terms of the nucleon state vector
\begin{equation}
\label{19a}
\left\langle p {1 \over 2} \left\vert
\int d^3x  \left(- g x^l {\bar \psi} B^{la} T^a \psi
+ x^l \partial ^s \left( E^{sa} B^{la} - E^{la} B^{sa} \right)
\right)  x^k \right\vert p {1 \over 2} \right\rangle  = 0
\end{equation}
then $L^i_{QCD} +G^i _{QCD}$ would be a proper angular momentum
density. Although 
these restrictions may seem {\it ad hoc} it should be noted that the
commutation relationships of $G^i _{QCD}$ with itself and with
other parts of the QCD angular momentum operator will in general
depend on the specific form of $E^{ia}$ and $B^{ia}$. Therefore
it would be more surprising if no restrictions were placed on
these fields by the commutation relationships, or if the commutation
relationships worked for any form of the fields. As an example one
can consider the fields of a color charge located at ${\vec R}$
and a color magnetic dipole, ${\vec m}$, located at the origin.
In the one-gluon exchange approximation \cite{don} the color 
fields will take the following electromagnetic-like form
\begin{equation}
\label{20}
E^{ia} = {g {x '}^i \over {r'}^3} T^a \nonumber \; \; \; \; \; \; \;
B^{ia} = \left( {3 x^i x^l m^l \over r^5} - {m^i \over r^3} \right)
T^a
\end{equation}
where ${\vec x}' = {\vec x} - {\vec R}$, the color dipole has been
oriented along the positive z-axis, and the right hand sides should be
bracketed by $\psi ^{\dag}$ and $\psi$. From Refs. \cite{d1} \cite{das}
this particular field configuration results in ${\vec L} _{QCD}
+{\vec G}_{QCD}$ satisfying the angular momentum commutator
({\it i.e.} the second term in Eq. (\ref{18}) does not contribute).
The condition in Eq. (\ref{19}) does not uniquely determine the 
color fields, since $E^{la} = B^{la}$ and $x^l B^{la} = 0$ also 
satisfy Eq. (\ref{19}). As another example one could postulate
that the color fields are split into a perturbative plus
non-perturbative part $\rightarrow Perturbative + Non-perturbative$.
The perturbative part could be of some form similar to the one-gluon
exchange form given in Eq. (\ref{20}), while the non-perturbative
part could be the purely chromoelectric flux tubes that are thought
to form between quarks in the standard picture of confinement. As
already mentioned a perturbative part of the form given in
Eq. (\ref{20}) would make the commutation relationships come
out correctly, and a purely chromoelectric ( {\it i.e.} $B^{la} = 0$),
non-perturbative part would also make the commutation relationships
come out correctly so that Eq. (\ref{19}) would be satisfied.
 
Even if the above examples are not
entirely realistic or do not give a unique restriction of the fields,
the point is that even more realistic
field configurations, determined dynamically from the QCD field
equations via lattice gauge theory or some other method, may have 
kinematical restrictions coming from the angular momentum commutators.

In contrast, for electromagnetic systems such as the
electric charge/magnetic charge or electric charge/magnetic dipole,
one knows the form of the fields from the outset.
In these cases one just needs to check that the
fields lead to the correct angular momentum commutation
relationship for the sum of the orbital plus field angular momentum.
However, if the form of the electric and magnetic fields
of these systems were not given analytically,
then also in these cases one would find some restrictions on the form
of the fields coming from the angular momentum algebra.
Actually the Dirac condition on magnetic charges -- $e g = 1/2$ --
can be viewed as a restriction on the electric charge/magnetic
charge system that arises from the angular momentum algebra
\cite{peskin}. 

{\flushleft {\big . \bf{4. DISCUSSION AND CONCLUSION}}} 

Using the gauge-invariant form of the QCD angular momentum operator
given in Ref. \cite{ji} we have investigated the commutators of
the various terms in Eq. (\ref{2}).   
There were two crucial assumptions made in arriving at
these results. First, we only considered up to one-loop
corrections to the various parts of the angular momentum
operators. This greatly simplified the analysis since
the quark spin operator (the first term in Eq. (\ref{2}))
only begins to experience renormalization scaling effects at
the two-loop level. This implies that the quark spin
operator still satisfies the angular momentum
commutation relationships up to one-loop. This in turn
implies that $J_{QCD} ^i -S^i$, and therefore
$L_{QCD}^i + G_{QCD} ^i$, should satisfy the angular
momentum commutation relationships up to one-loop.
This requirement was the starting point in the
calculations leading to Eqs. (\ref{19}) (\ref{19a}).
Second, in order to obtain the commutator of some of the
angular momentum parts ({\it e.g.} the field part, $G_{QCD} ^i$)
we assumed that since the nucleon is a bound state that
the gauge potentials were time-independent so that we could
set $\partial _0 A_{\mu} ^a =0$ (or more rigorously
taking the time-independent condition to apply to the
state vector as $\partial _0 A_{\mu} ^a
\vert p \frac{1}{2} \rangle =0$). From Eq. (\ref{ccr}) the only
non-trivial field commutators involved $\partial _0 A_{\mu} ^a$
so for the nucleon the chromoelectric and chromomagnetic
field operators commuted with one another, which in turn
implied that $[G_{QCD}^i , G_{QCD}^j] = 0$. Without this
assumption both $[G_{QCD}^i , G_{QCD}^j]$ and the
commutators like $[G_{QCD}^i , L_{QCD}^j]$ would have been
much more complicated, and our final results --
Eqs. (\ref{19}) (\ref{19a}) -- would be substantially altered.

In analogy with certain electromagnetic systems, such
as the charge/monopole system \cite{peskin} or the charge/magnetic 
dipole system \cite{das}, where it is only the combination of 
orbital plus field angular momentum which satisfies the correct
commutation relationships, we examined the commutator for the
combination $L^i_{QCD} + G^i_{QCD}$. Although for $L^i _{QCD}$
we did not make any assumptions about the specific form
of the color fields, we did have to assume some general
form - Eq. (\ref{10}) - for the color fields in order to calculate the
commutators containing $G^i _{QCD}$. The calculation of
the commutator of $L^i_{QCD} + G^i_{QCD}$, given in Eq. (\ref{18}),
shows that in addition to the expected $i \epsilon ^{ijk} (L^k _{QCD}
+ G^k _{QCD})$ there were additional terms (the second line on the
right hand side of Eq. (\ref{18})) which ruined the closure of the
commutation relationship. We took this to indicate one of
the following two possibilities :
\begin{enumerate}
\item
The assumptions about the particular form of the color fields
({\it i.e.} Eq. (\ref{10}) or that the color fields are static)
were not valid. However, given the close
connection of this form of the fields with the basic structure
of Yang-Mills theory it is not apparent how one could alter this
with out altering the structure of Yang-Mills theory.
\item
The form of the color fields are restricted in some way
({\it e.g.} Eqs. (\ref{19}) or (\ref{19a})) so that
$L^i _{QCD} + G^i _{QCD}$
does satisfy the angular momentum commutation rules. 
\end{enumerate}
Both of these possibilities indicate that some restrictions are
placed on the form of the color electric and magnetic fields by the
commutation relationships. 

{\flushleft {\big . \bf{5. ACKNOWLEDGMENTS}}} 

This work has been funded by the
National Research Council under the Collaboration in Basic Science and
Engineering Program. VD is supported by a Georg Forster Research
Fellowship by the Alexander von Humboldt
Foundation. The authors would like to thank
Xiangdong Ji for valuable discussions about this work.

\end{document}